\documentclass[12pt]{iopart}

\usepackage{iopams}
\usepackage{graphics}

\newcommand{\s}{$s$-shell}
\newcommand{\p}{$p$-shell}
\newcommand{\figwidth}{8.5cm}

\begin{document}

\title[$p$-shell hybridisation and {H}und's rules
mitigation]{$\boldsymbol{p}$-shell hybridisation and {H}und's rules
  mitigation:\\
  Engineering Hilbert spaces in artificial atoms}

\author{Jordan Kyriakidis\footnote{URL: http://soliton.phys.dal.ca}}

\address{Department of Physics and Atmospheric Science, Dalhousie
  University, Halifax, Nova Scotia, Canada, B3H 3J5}

\ead{jordan.kyriakidis@dal.ca}

\date{\today}

\begin{abstract}
  The magnetic-field dependence of many-body states in quantum dots
  can be tailored by controlling the mixing of various angular
  momenta.  In lateral quantum dots---defined electrostatically in a
  two-dimensional electron gas---this mixing can be accomplished by
  introducing anisotropies in the confinement potential, thereby
  explicitly breaking rotational symmetry.  Mixing can be severe
  enough to violate Hund's rules, even at zero magnetic field.  We
  illustrate the principle through calculations of states and spectra
  of four-electron droplets ($p$-shell) with long-range Coulomb
  repulsions and confined in anisotropic potentials.  Our results show
  that the Hilbert space in these nanostructures can be engineered to
  the particular application domain.
\end{abstract}



\section{Introduction}
\label{sec:introduction}

In order for quantum nanoelectronics to become a demonstrably viable
technology, the particular nature of the discrete states of a finite
system---a quantum dot, for example---must be both known and
controlled.  This is also a natural prerequisite for controlling the
coherent time evolution of a nanoscopic quantum system.

In the present work, we study the specifics of particular quantum
dots~\cite{ciorg00:addit.spect.later} defined electrostatically within
a two-dimensional electron system (lateral quantum dots).  By
adjusting the shape of the confinement potential, individual many-body
states within the quantum dots, and within the leads, may be tailored
to exhibit a desirable magnetic-field dependence.  (See, for example,
Ref.~\cite{fuhrer01:trans.steep.walls}.)  Among similar lateral
structures, the present dots dots have the distinguishing feature that
the number of electrons can be controlled down to a single
electron~\cite{ciorg00:addit.spect.later}, similar to vertical
devices~\cite{kouwen01:few.elect.quant.dots,matag02:self.consis.simul}.
They differ from vertical devices in that the confinement potential
can be generally controlled to a greater degree; in the dots we study
here, confinement is produced by four different gates.  Because of
this, singlet-triplet transitions~\cite{kyriak02:voltag.tunab.singl}
and negative differential
resistance~\cite{ciorg02:tunab.negat.differ}, for example, can be
tuned at a \emph{fixed} magnetic field by adjusting the gate voltage
alone.  Further, they differ from vertical devices in that the leads
of lateral devices are the edges of a high-mobility two-dimensional
electron gas (2DEG).  In such devices, spin-splitting of the edge
states, due primarily to exchange effects, results in a spatial
separation of the spin channels.  Consequently, the tunnelling rates
into and out of the quantum dot are measurably different for each spin
species.  This additional diagnostic tool has been recently
used~\cite{ciorg02:collap.spin.singl} to show how interactions force
the collapse of the spin-singlet state at filling-factor $\nu=2$.

The present lateral dots have a relatively small confinement energy
($\omega_0 = 1$~meV) compared to, say, vertical devices.  Because of
this, the system is more strongly interacting (Coulomb
interactions~$\sim \omega_0^{1/2}$, kinetic energy~$\sim \omega_0$).
This enhanced interaction strength largely drives the various phases
we describe below.  The dots are also
anisotropic~\cite{reiman02:elect.struc.quant.dots,szafr04:anisot.quant.dots,szafr04:re.entran.pinnin}
and, consequently, the many-body states of the system contain mixed
angular momenta.  By studying the magnetic-field evolution of these
states, their angular-momentum composition can be deduced.

The dots are well-isolated from the leads and we work in the Coulomb
blockade regime; transport proceeds via single-particle sequential
tunnelling into and out of the quantum dot.  We thus assume an isolated
dot and look at its intrinsic properties.

\section{The isotropic regime}
\label{sec:isotropic-case}

Insight can be gained by looking first at the simple case of parabolic
confinement.  We first review the relevant spectra for dots containing
four electrons.  In these two-dimensional dots, the \s\ has one
orbital with angular momentum $L_z = 0$ and the \p\ has two degenerate
orbitals with angular-momentum $L_z = \pm 1$.  Application of a
magnetic field perpendicular to the 2DEG breaks the orbital
degeneracy.

The noninteracting Hamiltonian is
\begin{equation}
  \label{eq:9}
  H = \frac{1}{2m^*} \left( \bi{p} + \frac{e}{c}\bi{A}(\bi{r})
    \right)^2 + \frac{1}{2}m^* \omega_0^2 r^2 +
    g\mu_B\bi{B}\cdot\bi{S},
\end{equation}
where $\bi{r} = (x, y)$ is the two-dimensional position operator.  The
single particle states are the well known~\cite{fock28,darwin30}
Fock-Darwin states, or alternatively~\cite{jacak97:quant.dots}, and
more useful in the present context, a set of two decoupled harmonic
oscillators $|mn\sigma\rangle$:
\begin{equation}
  \label{eq:1}
  E_{mn\sigma} = \hbar\Omega_+\left(n+\frac{1}{2}\right) + 
  \hbar\Omega_-\left(m+\frac{1}{2}\right) + g\mu_BB\sigma.
\end{equation}
In this notation, the $z$-component of angular momentum of the state
$|mn\sigma\rangle$ is $L_z=\hbar(n-m)$, $n$ is the Landau-level index,
and
\begin{equation}
  \label{eq:2}
  \Omega_{\pm} = \frac{1}{2} \left(
    \sqrt{\omega_c^2 + 4\omega_0^2} \pm \omega_c \right),
\end{equation}
with $\omega_c=eB/(m^*c)$ the cyclotron frequency and $\omega_0$ the
parabolic confinement frequency.  We also take the magnetic field
$\bi{B}$ in \eref{eq:9}, and throughout this paper, to be
directed perpendicular to the two-dimensional plane of the dot.  The
final term in \eref{eq:1} is the Zeeman energy.

When more than one electron is trapped in the dot, long-range Coulomb
repulsion plays a significant role, particularly for few-electron
droplets, where screening is minimal and the long-range nature of the
interaction is manifest.  We take the full Coulomb interaction $V_C =
e^2/(\varepsilon |\bi{r}_1 - \bi{r}_2|)$ and solve the problem
numerically by exact diagonalisation, exploiting the spin and
rotational invariance of the model.  (See
Refs.~\cite{wensauer04:config.inter.method.fock,kyriak02:voltag.tunab.singl}
for details.)

\begin{figure}
  \centering
  \resizebox{\figwidth}{!}{\includegraphics*{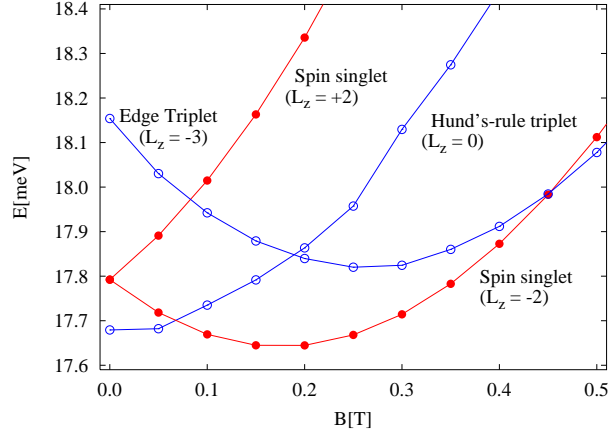}}
  \caption{Ground-state transitions for 2D parabolic dots containing
    four electrons with confinement $\omega_0 = 1$~meV.  The figure
    shows those low-lying states which will be significantly mixed by
    the anisotropy.  (See text.)  The states shown all have $S_z=0$
    and material parameters are for GaAs.}
  \label{fig:N1-4}
\end{figure}

Figure~\ref{fig:N1-4} shows the successive ground states for four
particles~\cite{reiman02:elect.struc.quant.dots}, for GaAs (material
parameters: dielectric constant $\varepsilon = 12.4$; effective mass
$m^* = 0.067$; $g$-factor $g = -0.44$).
At zero field, the ground state configuration is a spin triplet with
zero angular momentum.  We label this state ``Hund's-rule state''
because the dominant component, approximately 75\%, of this correlated
state is the Hund's-rule state $|00\downarrow, 00\uparrow,
10\downarrow, 01\downarrow\rangle$~\footnote{Our notation
  $|\cdots\rangle$ denotes appropriately antisymmetrised and
  normalised states.  Additionally, the one-body state
  $|10\downarrow\rangle$, for example, denotes the Fock-Darwin state
  with orbital quantum numbers $m=1$ and $n=0$, and spin projection
  $S_z=-\hbar/2$.}.  The first crossing occurs already below 0.1~T,
and the new ground state is a spin singlet state with angular momentum
$L_z = -2$.  The dominant component of this state is the usual $\nu =
2$ state, $|00\downarrow, 00\uparrow, 10\downarrow,
10\uparrow\rangle$.  Hence, this transition involves both a spin and
an orbital transition.  The second ground-state transition of
relevance here, occurring at approximately 0.45~T, is to an ``edge
triplet'' state.  This is again a spin and orbital transition; the new
ground state is a spin triplet and has angular momentum $L_z = -3$.
The dominant contribution to this state is the single Slater
determinant $|00\downarrow, 00\uparrow, 10\downarrow,
20\downarrow\rangle$.  We term this state an ``edge triplet'' since
the triplet is formed between a $p$ orbital and the $L_z = -2$
$d$-shell orbital, unlike the Hund's-rule state whose triplet is
formed by the two $p$ orbitals.  (The $d$-shell orbital, having
greater diameter, is closer to the edge of the droplet than the
$p$-shell.)  Continuing the magnetic field evolution would lead
eventually to the $\nu=1$ state, followed by fractional occupation of
the lowest-Landau-level orbitals.

Just below 0.2~T, there is a transition between two \emph{excited}
states (see Fig.~\ref{fig:N1-4}), both of which are triplets; one is
the Hund's-rule state, the other the edge triplet.  Explicitly
breaking the rotational (spatial) symmetry allows mixing between
different angular-momentum states, but not different spin states.  The
edge triplet and Hund's rule triplet are both strongly mixed near the
transition.

\section{Anisotropy-induced mixing}
\label{sec:inter-anis-dots}

The nature of the anisotropy in the confinement potential of the
present dots has been explored in
Ref.~\cite{kyriak02:voltag.tunab.singl}, where the confinement
potential has been found to fit very well the analytic form
\begin{equation}
  \label{eq:3}
  V_{\mathrm{con}} = \frac{1}{2}m^* \omega_0^2 \left[
    \left(x^2+y^2\right) +
    \gamma\left(x - \frac{y^2}{D}\right)^2\right].
\end{equation}
The gate geometry producing such confinement is sketched in the inset
of Fig.~\ref{fig:hund}.  For the four-electron droplet, we find
$\gamma=0.5$ and $D=5a_0$, where $a_0=\hbar^2/(m^*e^2)$ is the
effective Bohr radius.  The first term in parentheses is the usual
parabolic confinement, and the second term in parentheses is the
explicit symmetry-breaking term which induces mixing in the angular
momentum states.  Because the anisotropy is of quartic order, non-zero
matrix elements exist between states with angular momentum $L_z$ and
$L_z \pm \delta R$, where $\delta R = 0,1,2,3,4$.  The respective
Hilbert spaces of different \emph{spin} angular momenta remain
segregated.

It is useful to consider the Hamiltonian as two terms
\begin{equation}
  \label{eq:4}
  H = H_0 + H_1.
\end{equation}
$H_0$ contains the full Coulomb interactions, the kinetic energy, and
the isotropic (parabolic) confinement potential, whereas $H_1$
contains only the anisotropic piece of the confining potential---the
term proportional to $\gamma$ in \eref{eq:3}.  We can label the
eigenstates of $H_0$ as $|i, L_z, S\rangle$, which denotes the $i$-th
state with $z$-component of angular momentum $L_z$, and spin $S$:
\begin{equation}
  \label{eq:5}
  H_0 |i, L_z, S\rangle = E_i^{(0)}(L_z, S) |i, L_z, S\rangle.
\end{equation}
(In practise, we also use $S_z$ as a good quantum number.)
Figure~\ref{fig:N1-4} shows the field-dependence of several
$E_i^{(0)}$.

The eigenstates $|j, S\rangle$ of the full anisotropic Hamiltonian $H$
in \eref{eq:4} do not preserve angular momentum.  These states
can be constructed from the parabolic states $|i, L_z, S\rangle$:
\begin{equation}
  \label{eq:6}
  |j, S\rangle = \sum_{i, L_z} \alpha_i^{L_z} |i, L_z, S\rangle.
\end{equation}
The probability $P(L_z)$ that a given state $|j, S\rangle$ has angular
momentum $L_z$ is then given by
\begin{equation}
  \label{eq:7}
  P(L_z) = \sum_i \left| \alpha_i^{L_z} \right|^2,
\end{equation}
with the sum rule $\sum_{L_z} P(L_z) = 1$ conserving probability.

Our calculations of four interacting electrons in an anisotropic
quantum dot, \eref{eq:4}, and \eref{eq:3}, proceed through two
successive exact-diagonalisation procedures.  In the first step, we
compute the energy levels and eigenstates of four interacting
electrons in a parabolic potential, $H_0$ in \eref{eq:4}.  As a basis,
we use antisymmetrised four-electron states of the 2D harmonic
oscillator (Fock-Darwin) states $|mn\sigma\rangle$.  We use the full
long-range Coulomb interaction, whose exact matrix elements are given
in Ref.~\cite{kyriak02:voltag.tunab.singl}.  We ensure the accuracy of
our results by progressively increasing the Hilbert space and
monitoring the convergence of the energy levels.  This yields the
$E_i^{(0)}(L_z, S)$ and the $|i, L_z, S \rangle$ of \eref{eq:5}.

In the second step, we diagonalise the full anisotropic Hamiltonian.
The basis in this step is formed from the correlated eigenstates, $|i,
L_z, S \rangle$, computed in the previous step.  The notation ``$|i,
L_z, S \rangle$'' indicates the $i$th eigenstate of $H_0$ (which
includes full Coulomb interactions) with angular momentum $L_z$ and
spin $S$.  Using this technique, the size of the Hilbert space and the
degree of angular-momentum mixing can be independently controlled, and
the convergence of the resulting energies and states can be readily
investigated.  Total spin $S$ and $z$-component $S_z$ are good quantum
numbers and are used to classify our states.

\subsection{Ground-state spin transition}
\label{sec:zero-field-regime}

\begin{figure}
  \centering \resizebox{\figwidth}{!}{\includegraphics*{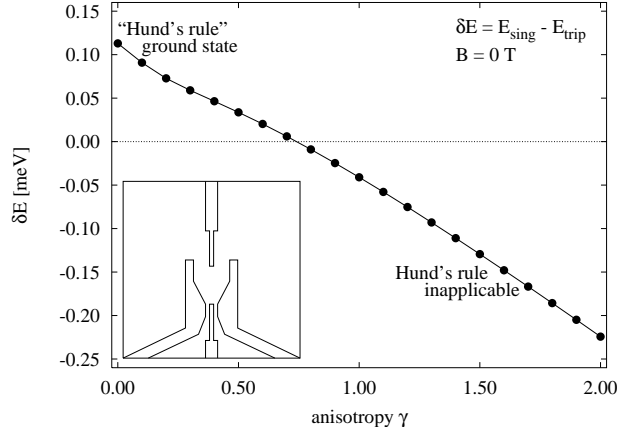}}
  \caption{Main plot: Energy difference $\delta E$ between the
    lowest-energy singlet and triplet states.  Positive (negative)
    $\delta E$ indicates a triplet (singlet) ground state.  The Hund's
    rule ground state exists only for $\delta E > 0$.  Inset: Sketch
    of the gate geometry producing anisotropic confinement,
    \eref{eq:3}.}
  \label{fig:hund}
\end{figure}

We first discuss the zero-field case.  Figure~\ref{fig:N1-4} shows a
triplet Hund's-rule ground state.  This state has a filled \s\ and two
singly-occupied $p$-orbitals.  In contrast, the first-excited state is
a singlet with a doubly-occupied $s$-orbital and one doubly-occupied
$p$-orbital ($L_z = -1$).  At zero field, the singlet is degenerate
with its time-reversed partner---a doubly-occupied $p$-orbital with
$L_z = +1$.  Because the singlet state is degenerate and the triplet
is not, we expect anisotropy to mix the singlet state more strongly
than the triplet.  As a consequence, the singlet-triplet gap should
decrease as a function of anisotropy.  (In other words, the
lowest-energy triplet should rise faster in energy with anisotropy
than the lowest-energy singlet.)  This is indeed the case, as shown in
Fig.~\ref{fig:hund}.
In essence, the anisotropy-induced \p\ hybridisation is strong enough
to produce two states delocalised across the $L_z = \pm 1$ orbitals.
The splitting between these hybridised states is sufficient to
overcome the exchange-energy savings associated with the triplet
state.  Hund's rules are thus mitigated and the spin of the ground
state can be tuned through the anisotropy.

\subsection{Field-dependence and angular-momentum content}
\label{sec:field-dependence}

\begin{figure}
  \centering
  \resizebox{\figwidth}{!}{\includegraphics*{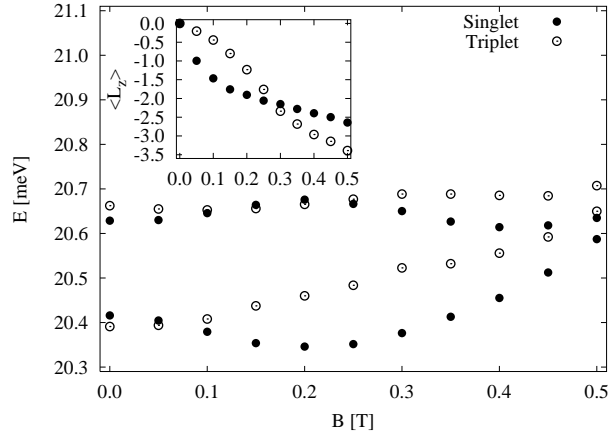}}
  \caption{Main plot: Filled circles indicate the two
    lowest-energy singlet states for the four-electron interacting and
    anisotropic droplet.  Open circles are the same, but for triplet
    states.  The figure is for anisotropy parameters $\gamma=0.5$ and
    $D=5a_0$, and for $\omega_0=1$~meV.  All other parameters are
    taken for GaAs.  Inset: Average angular momentum $\langle L_z
    \rangle$ for the lowest-energy singlet and triplet states for the
    same set of parameters, as a function of magnetic field $B$~[T].}
  \label{fig:n4spectrum}
\end{figure}

In addition to the ground-state spin, the entire field-dependence of
the many-body states can be tuned.  The states analogous to those in
Fig.~\ref{fig:N1-4} are shown in Fig.~\ref{fig:n4spectrum}.
The relatively weak field dependence of these states, is an indication
of the angular-momentum mixing induced by the anisotropy.  (Compare
with Fig.~\ref{fig:N1-4} which shows states with definite angular
momentum for the same energy interval (0.8~meV).)

The average angular momentum
\begin{equation}
  \label{eq:8}
  \langle L_z \rangle \equiv \langle j, S | L_z | j, S \rangle
\end{equation}
of the states shown in Fig.~\ref{fig:n4spectrum} are not constants but
are themselves dependent on the magnetic field.  This field dependence
is shown in the inset of Fig.~\ref{fig:n4spectrum}.  At zero magnetic
field, both states have $\langle L_z \rangle =0$.  This is a
reflection of time-reversal symmetry, whose consequence is that the
energy of each angular-momentum state depends on the magnitude, not
the sign, of the angular momentum.  Hence, $+L_z$ and $-L_z$ states
are mixed equally.  This degeneracy is lifted when time-reversal
symmetry is explicitly broken, in favour of the negative
angular-momentum states.

\begin{figure}
  \centering \resizebox{\figwidth}{!}{\includegraphics*{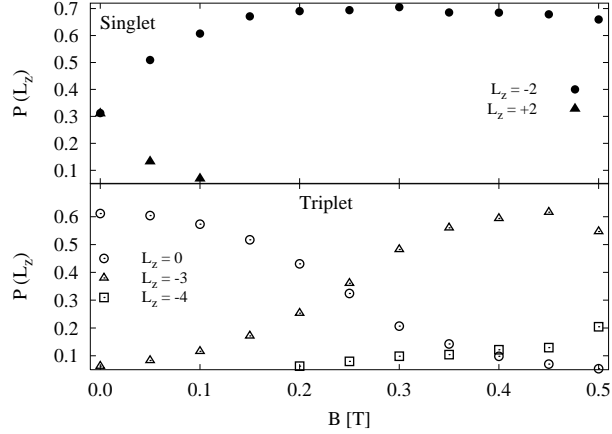}}
  \caption{Top: Dominant angular-momentum composition of the
    lowest-energy singlet state for the four-electron droplet.  The
    figure is for anisotropy parameters $\gamma=0.5$ and $D=5a_0$, and
    for $\omega_0 = 1$~meV.  the mixing occurs primarily at zero
    field.  Many more angular-momentum states contribute to this
    state, but these all have less than 10\% contribution.
    Bottom:~Same plot, but for the lowest-energy triplet.  In this
    case, the mixing occurs primarily at moderate (0.25~T) fields.}
  \label{fig:n4lz}
\end{figure}

We focus first on the singlet.  In the isotropic case, the
lowest-energy singlet has $L_z=-2$.  (See Fig.~\ref{fig:N1-4}.)  With
the anisotropy present, the inset of Fig.~\ref{fig:n4spectrum} shows
that $\langle L_z \rangle$ drops quickly from its zero-field value of
$\langle L_z \rangle =0$, and then, above 0.1~T, the field-dependence
is much weaker, ranging from about $\langle L_z \rangle = -2$ to
$-2.5$.  This indicates that the singlet state is primarily composed
of the isotropic ($L_z=-2$) singlet state above a field of
approximately 0.1~T.  The angular-momentum mixing is more severe below
this field value.  These results are born out by further analysis.  In
Fig.~\ref{fig:n4lz}, we plot the individual angular-momenta
contribution, \eref{eq:7}, for the singlet state.  Only those
angular momenta with $P(L_z) \ge 0.1$ are shown~\footnote{The
  calculations were performed by mixing 31 different angular-momentum
  sectors, $L_z=-20$ to $L_z=+10$.}.  Figure~\ref{fig:n4lz} shows
that, above 0.1~T, the isotropic $L_z=-2$ state is by far the dominant
contribution; the contribution of its time-reversed partner, $L_z=+2$,
dies out rapidly.  Thus, the anisotropy does not severely mix angular
momenta for this state above 0.1~T.

Figure~\ref{fig:n4spectrum} shows that the situation is rather
different for the triplet state.  Here, $\langle L_z \rangle$
decreases with magnetic field at a more uniform rate than does the
singlet.  In this case, significant mixing occurs between the edge
triplet and the Hund's-rule state, also a triplet.  (See
Fig.~\ref{fig:N1-4}.)  In Fig.~\ref{fig:n4lz}, we show the dominant
angular momentum contributions to this lowest-energy triplet state,
$P(L_z) \ge 0.1$.  We see that the makeup of this state changes as a
function of magnetic field, being composed primarily of the
Hund's-rule state at low fields, and the edge triplet at fields above
about 0.25~T.  Another state, with $L_z = -4$, emerges above the
background for fields greater than 0.3~T.  Thus, the anisotropy
induces significant angular-momentum mixing for the triplet state,
although not at zero field.

\section{Summary}
\label{sec:summary}

Well-isolated quantum dots containing a known and small number of
interacting electrons are simple enough to be amenable to a
theoretical treatment through directly diagonalising the relevant
many-body Hamiltonian.  By employing the successive
exact-diagonalisation procedures outlined above, calculations can be
extended with reasonable efficiency to anisotropic systems with lower
symmetry.  At the same time, these systems are amenable to
experimental analysis as well.  With high source-drain spectroscopy,
excited states as well as ground states can be directly imaged in
experiment.

In the present work, we analysed ground and excited states in a
four-electron lateral quantum dot.  We find that the (many-body)
long-range Coulomb repulsion and the (single-body) anisotropic
confinement potential both play important roles and both must be
included in any analysis.  By analysing the magnetic-field dependence
of these many-body levels, we have determined the average angular
momentum of each level, which angular-momentum states are mixed
together, how much they are mixed together, for both the ground and
excited states of the system.

The interplay between experiment and theory can be exploited to design
gate geometries which in turn produce confinement potentials whose
magnetic field dependence can be predetermined to be as desired---a
flat field-dependence, for example.  These are the elements necessary
to achieve quantum-state engineering, and a necessary prerequisite for
controlling, in real time, the coherent time evolution of individual
states that are robust toward decohering influences.

\begin{ack}
  The author acknowledges discussions with Pawel Hawrylak.  He is also
  grateful for the provision of unpublished experimental data from
  Andy Sachrajda.  This work was supported by NSERC of Canada and by
  the Nanoelectronics Program of the CIAR.
\end{ack}


\end{document}